# Efficient charge transfer in solution-processed PbS Quantum Dot-reduced graphene oxide hybrid materials

Beatriz Martín-García,[a,b] Anatolii Polovitsyn,[a,c] Mirko Prato[a] and Iwan Moreels[a,b]*

a. NanoChemistry Department, Istituto Italiano di Tecnologia, Via Morego 30, IT-16163 Genova, Italy. E-mail: Iwan.Moreels@iit.it

b. Graphene Labs, Istituto Italiano di Tecnologia, Via Morego 30, IT-16163 Genova, Italy.

c. Department of Physics, University of Genoa, Via Dodecaneso 33, IT-16146 Genova, Italy.

**Abstract** - Quantum dot - graphene hybrid materials have raised significant interest due to the unique synergy of the optical properties of colloidal quantum dots (QDs) and the transport properties of graphene. This stimulated the development of low-cost and up-scalable solution-processed strategies for hybrid materials with potential application in light harvesting and opto-electronic devices. Here we report a versatile covalent-linking based approach for the functionalization of reduced graphene oxide (rGO), to prepare a variety of QD-rGO hybrid dispersions with QDs of different size and composition (PbS, PbS/CdS and CdSe QDs), and shape (CdSe/CdS dot-in-rods). We achieved a well-controlled QD coverage of the rGO sheets by functionalizing the rGO surface with mercapto-silane linkers. A further spectroscopic investigation of near-infrared PbS QD-rGO materials demonstrates efficient electronic coupling between both materials. The QD photoluminescence emission quenching and exciton lifetime shortening up to 95%, together with subtle graphene Raman G-band shifts upon QD linking, supports electron transfer as the dominant relaxation pathway from the QD to the rGO. The use of core/shell PbS/CdS QDs allows tuning of the transfer efficiency from 94% for a 0.2 nm thin CdS shell, down to 30% for a 1.1 nm thick shell.

## Introduction

A continuing search for solution-processed materials that are low-cost and can be fabricated and applied on large scale is paramount to drive the development of future photonic devices such as solar cells,[1-3] photodetectors,[3-6] LEDs[7] or lasers.[8] Efforts are focused on both the optically active material as well as charge transport layers. In this respect, since the discovery of the extraordinary electronic properties of graphene,[9] it has often been incorporated as a nearly transparent and flexible electrode. In solution-based approaches, it is convenient to use graphene derivatives such as liquid-phase exfoliated graphene or reduced graphene-oxide (rGO), with rGO offering advantages such as large area coverage as well as tunable opto-electronic properties due to the presence of oxygen functionalities, which also favor chemical interactions.[5,10-14] While the light-harvesting properties of graphene or rGO could in principle also be exploited, typically different strategies are followed, for instance interfacing them with other materials such as colloidal quantum dots (QDs).[15-19] These possess a strong and tunable light absorption and emission together with a high chemical and photophysical stability, and therefore make an excellent photoactive material.[20,21] Additionally, combining them with graphene-based materials would at the same time resolve a major shortcoming of nanosized quantum dots, as QD thin films are known to suffer from inefficient carrier transport.[22,23] Thus, the potential of the resulting hybrid material clearly lies in the unique synergy of the optical properties of the QDs and the transport properties of the graphene.[12,24,25]

Compared to non-interacting composite systems, a clear increase of device photo-conversion efficiency and photo-catalytic activity has been demonstrated when a covalent linking is promoted between the optically active and charge transport materials.[16,17,26] Among the existing approaches to prepare such hybrid materials, the most common are based on a one-pot in situ synthesis, layer-by-layer electrostatic assembly, or click-chemistry using amino-based functional linkers.[17,18,26-28] Especially the latter is advantageous, allowing to first optimize each constituent individually, followed by coupling with precisely controlled distance. Following this principle, here we developed a covalent linking method that efficiently anchors a variety of QDs with different chemical composition, size and shape to rGO using short-chained silane molecules. By functionalizing rGO with (3-mercaptopropyl) trimethoxysilane (MPTS), we achieved a uniform and density-controlled QD coverage of the rGO sheets. To demonstrate the potential of the QD-rGO coupling approach, we used near infrared (NIR)-emitting PbS QDs to investigate the opto-electronic properties of PbS QD-rGO hybrid materials and the charge/energy transfer at their interface. Several groups have already focused on the understanding of energy and electron transfer processes in different Cd-based QDs and GO-or graphene-based systems.[24, 29-32] In contrast, PbS QDs are less studied despite their potential to significantly open up the spectral range of applications due to their narrow bulk band gap of 0.4 eV. Here we show that the proposed coupling leads to a highly efficient carrier transfer between the materials, reaching up to 95% as determined from the PbS QD photoluminescence (PL) quenching and exciton lifetime shortening. Moreover, by using a 0.2 nm - 1.1 nm CdS shell synthesized via cation exchange, we were able to carefully tune the charge transfer efficiency. The shell thickness dependence, and shifts in the rGO Raman G-band after functionalization and QD coupling suggest an electron transfer mechanism as the main pathway for carrier relaxation into the rGO, most likely favored by the short (< 1 nm) coupling distance in our QD/rGO hybrid materials.

**Results and discussion**

To build devices such as photodetectors or solar cells, two materials are necessary: an absorber, here QDs, and a transport layer, here chemically reduced graphene oxide (ESI, Scheme S1 and Fig. S1). The GO reduction has been carried out by caffeic acid, enabling the use of efficient green alternatives and at the same time avoiding residual functionalization by other reducing agents such as hydrazine.[29, 33] The degree of reduction is determined via X-ray photoelectron spectroscopy (XPS), with an O/C ratio of 0.40 for the initial GO, that is reduced to 0.22 for rGO, comparable to recent results for green reducing agents (0.14), while having reduced three-fold the reaction time.[14]

**Table 1.** Atomic ratio of the different elements in GO and rGO, and the corresponding functionalized derivatives, as determined with XPS using the amount of carbon as a reference.

| Sample | O/C | N/C | S/C | Si/C |
|---|---|---|---|---|
| **GO** | 0.40 | - | - | - |
| Silane-*f*-GO | 0.46 | - | 0.12 | 0.17 |
| **rGO** | 0.22 | - | - | - |
| ABA-*f*-rGO | 0.27 | 0.02 | - | - |
| ATP-*f*-rGO | 0.25 | 0.02 | 0.02 | - |
| Silane-*f*-rGO | 0.37 | - | 0.10 | 0.16 |

To attach QDs to rGO, common click-chemistry approaches using aminobenzene derivates such as 4-aminobenzoic acid (ABA) and 4-aminothiophenol (ATP) have been already proposed.[34, 35] However, those routes led to a low degree of functionalization and non-controlled QD/rGO assembly (see ESI for further details and results, Fig.s S2-S3). Therefore, here we propose a novel approach using short-chained silanes, more specifically MPTS. This linker can bind to both rGO (*via* Si-O) as well as QDs (*via* SH). The coupling results from a simple mixing of MPTS with rGO in EtOH at 60 °C. Evaluating the S/C and Si/C atomic ratios by XPS, we obtained an average ratio of 0.13 (Table 1), significantly increasing the functionalization degree compared to ABA or ATP where we only achieved an N/C ratio of 0.02 (Table 1, see also ESI). This is probably due to the ability of MPTS to bind to the carboxylic acid at the sheet edges, as well as the rGO surface itself *via* an Si-O-C reaction through the methoxy groups of the silane and the epoxy groups of the rGO,[36-40] which are found throughout the carbon network (see ESI). A more detailed analysis of the $S_{2p}$ XPS spectrum in silane-functionalized rGO (silane-*f*-rGO) is shown in Fig. 1. The best fit for the $S_{2p}$ profiles is obtained by using two components. A minor contribution centered at 164.2 ± 0.2 eV, amounting to 2 ± 2 % of the total amount of sulfur in the silane-*f*-rGO samples can be related to S-S bonds,[41] arising from a small amount of coupling between MPTS linkers. However, the main component centered at 163.2 ± 0.2 eV, constituting 98 ± 2 %, can be assigned to unbound thiol moieties (-SH).[42] This confirms that the chemical interaction between MPTS and rGO is directed by the Si-O-C path, leaving the SH group exposed to bind to the QDs.

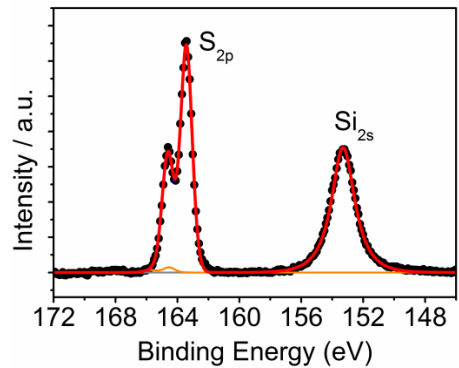

**Fig. 1.** XPS core-level $S_{2p}$ and $Si_{2s}$ spectra of silane-*f*-rGO. The $Si_{2s}$ peak is centered at (153.2±0.2) eV, a typical position for silanes.[38] For the $S_{2p}$ spectrum, the fitted curves correspond to unbound thiols (blue line) and a small amount of disulfide (orange line).

To promote the coupling between QDs and silane-*f*-rGO, we prepared the hybrid material in a mixed medium of toluene and EtOH at room temperature. We added OA-capped PbS QDs, dispersable in non-polar solvents (here toluene with polarity of 0.099[43]), to the silane-*f*-rGO dispersable in polar solvents (here EtOH with polarity of 0.654[43]), leading to the formation of aggregates over the course of 5 min. The resulting hybrid material then is precipitated by centrifugation and finally redispersed in DMF, a solvent with intermediate polarity (0.386[43]) in which a stable dispersion can be formed (for at least one month). Fig. 2a-c show TEM images for a typical hybrid material, using 5.0 nm PbS QDs. They reveal that the PbS QDs cover exclusively the rGO sheets, as no unbound PbS QDs are observed. Additionally, Fig. 2a shows how the PbS QD-decorated rGO sheets can even fold or overlap without affecting the QD-rGO coupling. Our approach also allows controlling the QD surface density by changing the relative PbS QD/rGO concentration in the starting solution (Fig. 3a-b). By increasing the QD/rGO ratio from 0.2 to 0.8 nmol QDs per µg of rGO it is possible to increase the density of QDs on the sheets by a factor of 1.7 (ESI, Fig. S4 and Table S1 for quantitative assessment). Interestingly, the degree of reduction also plays a role. Control experiments with GO, following the same MPTS functionalization and hybrid preparation procedure (see ESI for XPS, Fig. S4, TEM images, Fig.S5, and QDs density analysis, Fig.S7 and Table S1), show that, despite similar degree of functionalization obtained for the MPTS linkers (Table 1), rGO sheets typically show a 3-fold higher density of coupled PbS QDs. Possibly, this is due to the reduced surface polarization of rGO facilitating the binding of apolar OA-capped PbS QDs.

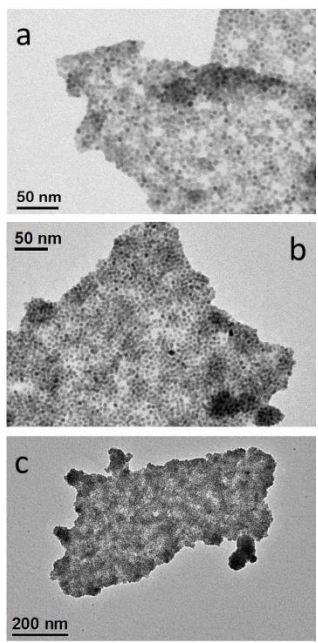

**Fig. 2.** Representative TEM images of core PbS QDs (5.0 nm diameter)/silane-*f*-rGO.

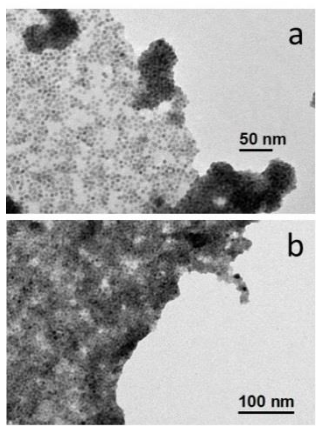

**Fig. 3.** Representative TEM images of PbS QDs (5.0 nm diameter)/rGO hybrid materials prepared from silane-*f*-rGO for the comparison of the QD coverage achieved varying the QD/rGO material ratio from 0.2 (a) to 0.8 (b) nmol QDs per µg rGO.

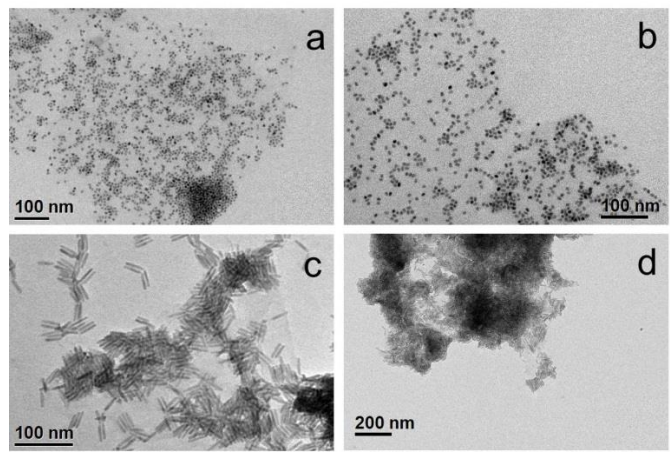

**Fig. 4.** Representative TEM images of hybrid materials prepared from silane-*f*-rGO with different materials: (a, b) CdSe QDs (5.8 nm), and (c, d) CdSe/CdS DIRs (28.3 nm length, 5.1 nm diameter).

The methodology to prepare hybrid materials from silane-*f*-rGO is also applicable to QDs of different composition and shape. As in the case of PbS QDs, spherical core CdSe QDs (Fig. 4a-b), and CdSe/CdS core/shell dot-in-rods (DIRs) (Fig. 4c-d) were found to exclusively cover the rGO, and were well dispersed over the sheets. Due to their elongated shape, the latter are linked along the DIR long axis, enabling a larger interfacial contact between the materials. This implies a high versatility of the proposed functionalization since, depending on the QDs one can target applications covering a wide spectral range from the visible to the NIR, while the QD-rGO interactions can be carefully tuned by adjusting the QD-rGO contact area in shape-controlled heterostructures.

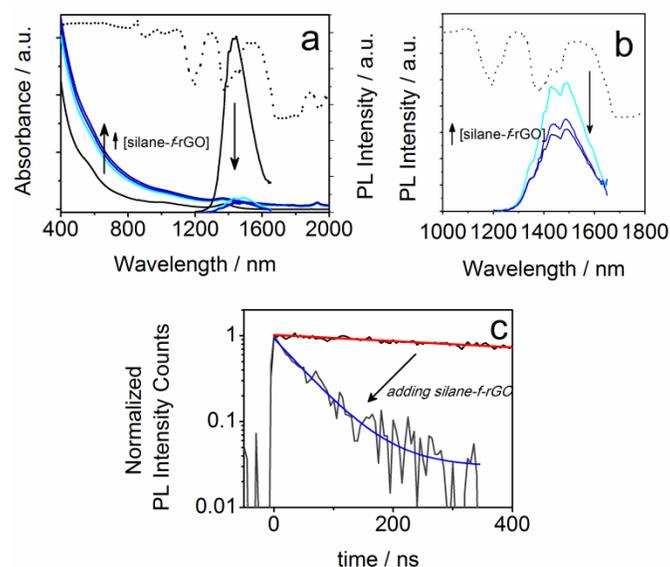

**Fig. 5.** Absorbance (a) and PL (a, b) spectra of PbS QD/silane-*f*-rGO hybrid materials dispersed in DMF with varying relative QD concentration (ratios 0.2, 0.4 and 0.8 nmol QDs per µg rGO). The data from the reference, OA-capped PbS QDs in TCE, are shown as black lines. (c) Corresponding PL decay traces and fits for pure QDs (red line) and rGO-hybrid materials (blue line), all prepared with a relative concentration of 0.2 nmol per µg rGO. The transmittance spectrum of DMF is shown as dotted line in (a, b).

To highlight the relevance of our solution-processable QD-rGO hybrid materials for photonic applications, we investigated the resulting optical properties. We focused our attention on PbS QDs, considering its importance for NIR solar cells, photodetectors, and LEDs,[1, 15, 23, 44-46] all of which strongly rely on efficient photon-electron conversion. In thin-film configuration they have already been combined showing an efficient energy or charge transport depending on the PbS QD- graphene distance.[15, 44, 47] Here, in our solution-based PbS QD/rGO hybrid material we also observed a significant PL quenching of up to 92% after coupling with silane-*f*-rGO (Fig. 5). This is accompanied by a PL lifetime reduction from 1.4 µs (PbS QDs in TCE) to merely 66 ns (95% reduction), among the highest values reported for QD/(r)GO systems.[29, 30] Possible effects of the preparation method on the PL quenching were discarded by various control experiments . For instance, the PL quenching is not simply due to the use of EtOH during the hybrid preparation, or the phase transfer of the QDs to DMF. Both were evaluated independently (see ESI, Fig. S8 and S9 and Table S2). EtOH washing only induced a severe PL quenching after three additional precipitation steps and left the PL decay unmodified otherwise. Additionally, we observed a somewhat reduced PL lifetime of 0.6 µs after transfer of 5.0 nm PbS QDs to DMF using 11-mercaptoundecanoic acid (MUA) ligands, however the retention of at least 40% of the PL decay time implies that in the QD-rGO hybrid materials, charge/energy transfer to rGO is the dominant pathway for the PL quenching. To quantify the effect of the coupling to rGO, we evaluated the transfer efficiency, $\eta$, via the effective decay rates of PbS QDs before, $k_{eff,QD}$, and after hybridization, $k_{eff,hybr}$, using OA-capped PbS QDs as a reference (Table 2).

$$\eta = \frac{k_{et}}{k_{et}+ k_{rad,QD}+k_{trap,QD}} = 1 - \frac{k_{eff,QD}}{k_{eff,hybr}} \qquad (1)$$

$k_{rad,QD}$ equals the QD radiative decay rate. This competes with two add channels: $k_{trap,QD}$, related to electron-hole trapping rates due to QD surface defects, and in the case of QD-rGO hybrids, $k_{et,graphene}$, which represents the energy/charge transfer rate from QD to rGO (Scheme 1). Although the transfer process and hybridization with rGO imply some uncertainty on $k_{trap}$ and thus $k_{eff,QD}$, control experiments with MUA capped PbS QDs in DMF suggest that these are minor (see ESI Fig. S8 and Table 2).

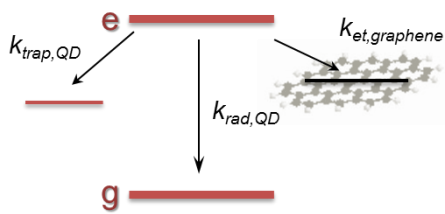

**Scheme 1.** Simplified diagram for excited state decay pathways in QD/rGO hybrid.

**Table 2.** PL lifetime, $\tau_{AV}$, and transfer efficiency, $\eta$ data obtained from the time-resolved measurements and Equation 1 (see text) for the different core PbS QD/silane-*f*-rGO hybrid materials prepared in DMF comparing with the starting PbS QDs in TCE.

| Sample | Solvent | $\tau_{av}$ / µs | $\eta$ |
|---|---|---|---|
| **OA-capped PbS QDs** | TCE | 1.4 | - |
| PbS QD/silane-*f*-rGO | DMF | 0.066 | 0.95 |

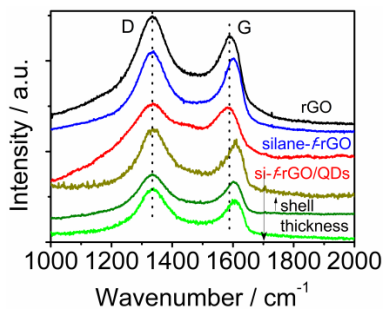

**Fig. 6.** Comparison between representative Raman spectra of the different PbS core and PbS/CdS core-shell QDs and silane-*f*-rGO hybrid samples. In the three lower spectra (green lines) we systematically varied the shell thickness. The Raman spectrum of the reference material rGO (black line) is also shown.

To improve our understanding on the nature of the energy/charge transfer processes in the hybrid system, we carried out complementary studies by means of Raman spectroscopy. This technique allows characterizing graphene-based materials by analyzing the position, relative intensity (ratio) and FWHM of the characteristic D and G bands. Specifically, the G band position shifts with physical and chemical changes in the carbon-network environment. Thus, it is highly sensitive to the presence of excess charges from, for instance, electron-donating molecules, leading to a downshift of the G band position, or electron-acceptor molecules promoting an upshift.[15, 30, 48] The silane functionalization of rGO leads to a G band upshift of 12 ± 2 cm$^{-1}$ (value determined from the average obtained on three different areas of the sample, see ESI Fig. S10), which agrees with a *p*-doping already observed by the introduction of different functional groups in the graphene network (Fig. 6).[38, 48] Meanwhile, the attachment of PbS QDs leads to a downshift of 15 ± 1 cm$^{-1}$ compared with the silane-*f*-rGO spectrum. In view of the *p*-doping after silane functionalization, this shift can be ascribed to an electron enrichment of the rGO network in PbS QD-rGO hybrid materials. Therefore, although energy and electron transfer pathways could still both be present, the short distance between the PbS QDs and the (r)GO sheets induced by the 0.7 nm MPTS linker promotes electron transfer as the main mechanism for the non-radiative relaxation of the QDs.[24, 29, 30, 47]

Considering the strong distance-dependence of the electron transfer, we were able to modulate the transfer efficiency by controlling the QD-rGO coupling length. We synthesized a thin CdS shell of 0.2 nm to 1.1 nm on the PbS QDs by cation exchange. This increases the effective QD-rGO distance, while leaving the assembly itself unaffected, again confirming the universal nature of the approach here developed (Fig. 7a-c). The resulting transfer efficiency, determined again from the PL lifetime, is displayed in Fig. 7d. We observed a notable decrease of the transfer efficiency upon increasing shell thickness (see also ESI, Table S3). As a control, the trend observed for rGO is confirmed by GO. However, we obtained consistently lower transfer efficiencies (this also agrees with a smaller relative shift of the Raman G band in GO-based hybrids, see ESI Fig.S11), demonstrating the importance of achieving a sufficiently high degree of reduction for rGO, not only to improve the assembly process but also to increase the transfer efficiency. The difference between GO and rGO likely lies in the modification of the graphene electronic structure.[24, 49] More specifically, the larger fraction of C $sp^2$ domains, increasing from 42% in the GO to 56% in the rGO (see ESI) leads to an increased electronic density of the states,[50] promoting more efficient charge transfer.

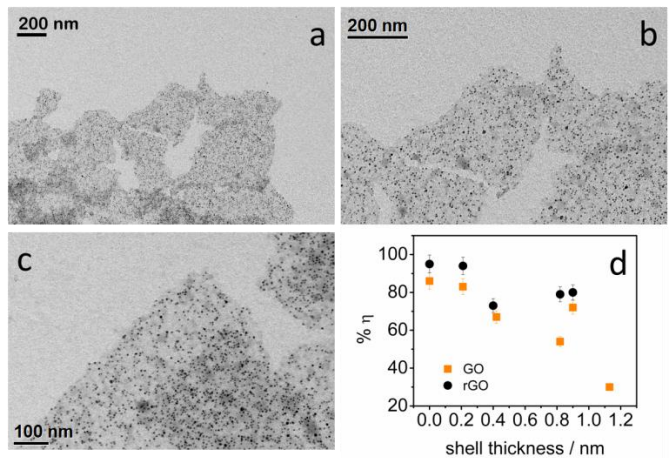

**Fig. 7.** (a-c) Representative TEM images at different scales of core-shell PbS/CdS QDs (core 4.2 nm diameter, shell 0.4 nm thickness) coupled to silane-*f*-rGO. (d) Influence of the CdS shell thickness on the transfer efficiency $\eta$ between the PbS QDs and the silane-*f*-(r)GO. Samples were prepared with a ratio of 0.2 nmol QDs per µg (r)GO.

By evaluating the electronic interaction between core/shell PbS/CdS QDs and rGO by means of Raman spectroscopy (Fig. 6), we confirm the role of the shell in inhibiting the electron transfer. When a thick shell is present the original G band position corresponding to the silane-*f*-rGO is recovered, indicating that the electron transfer between both materials is suppressed. Most importantly, both PL and Raman spectroscopy data demonstrate that it is possible to maintain a high electron transfer efficiency in PbS/CdS for thin shells of about 0.2 nm. This implies that such core/shell QDs can be for instance used in the implementation of solar cells or other applications operating under ambient conditions, as the CdS shell will improve the QD chemical and environmental stability.[45, 51]

**Experimental**

**Materials.** $PbCl_2$ (99.999%) and sulfur (99.999%) were purchased from Alfa-Aesar; CdO (≥99.99%) and trioctylphosphine oxide (TOPO, ≥90%) from Strem Chemicals; OA (90%), 1-octadecene (ODE, 90%), octadecylphosphonic acid (ODPA, 97%), hexylphosphonic acid (HPA, 95%), toluene (≥99.7%), ethanol (≥99.8%, without additive), tetrachloroethylene (TCE, anhydrous, ≥99%) and dimethylformamide (DMF, anhydrous, ≥99.8%) from Sigma Aldrich; oleylamine (OlAm, 80%) from Acros Organics. Graphite flakes were supplied by Graphene Supermarket; $H_2SO_4$ (95-98%), $NaNO_3$ (≥99.5%), $KMnO_4$ (≥99%, low in mercury), 4-aminobenzoic acid (ABA, ≥99%), 4-aminothiophenol (ATP, 97%), 3-(mercaptopropyl)trimethoxysilane (silane, 95%), caffeic acid (CA, ≥95%) and ammonium hydroxide solution (28%w $NH_3$ in $H_2O$, ≥99.99%) were supplied by Sigma Aldrich. The water used in the experiments was obtained from a MilliQ® system.

**Nanocrystal synthesis.** Oleic acid-capped PbS QDs were prepared following a slight modification of the method described in Moreels *et al.*[52] Briefly, for the synthesis 1 g of $PbCl_2$ and 7.5 mL of OlAm are degassed for 30 min at 125°C in a three-neck flask under Argon. Then, the temperature is adjusted to 120 – 160°C, depending on the desired QD size, and 2.25 mL of a 0.3 M OlAm-sulfur stock solution is added. After synthesis the OlAm ligands are replaced by adding OA to the QD suspension in toluene, followed by precipitation by EtOH and resuspension in toluene. The diameter is determined from absorption.[52] The CdS coating was carried out by cation exchange.[53] An OA-capped PbS QD solution of about 10 µM in toluene/ODE (1:1 vol.) was degassed for 1h under Argon at 80 ºC. Then, the solution was heated up to 100 ºC and 0.5 mL of a 0.3 M Cd-oleate stock solution was added to form the CdS shell. The reaction continued for 1 to 180 min depending on the desired shell thickness (0.2 - 1.1 nm). The shell thickness is determined from the blue shift of the first absorption peak after cation exchange.[53]

CdSe QDs and CdSe/CdS DIRs were prepared in accordance with a procedure published by Carbone *et al.*[54] Briefly, the CdSe QD synthesis was carried out mixing a clear solution formed by TOPO/ODPA/CdO with the injection of a Se:TOP solution at 370°C under nitrogen. The CdSe diameter and concentration were determined from the first absorption peak and the absorbance at 350 nm.[55] The CdSe/CdS DIRs were prepared *via* seeded growth by adding the CdSe seeds together with the sulfur precursor to a flask that contains a TOPO/ODPA/HPA/CdO solution at 350°C. The rod dimensions and the concentration were determined from TEM and elemental analysis by inductively coupled plasma optical emission spectroscopy (ICP-OES) on digested solutions, respectively.

**Preparation of graphene oxide and reduced graphene oxide.** We obtained the GO by a slight modification of the Hummers' method[56] from graphite flakes. A paste was formed by mixing graphite/$H_2SO_4$/$NaNO_3$/$KMnO_4$, after which the graphite was oxidized for 15 h for an improved exfoliation and oxidation of the material. To chemically reduce the GO, we used caffeic acid (CA) and introduced several modifications to the reaction in aqueous medium at 95ºC proposed by Bo *et al.*[14] Briefly, in the reduction of a 0.1 mg mL$^{-1}$ GO aqueous dispersion we have achieved a similar degree of reduction with a smaller amount of CA (CA:GO weight ratio of 1:1) and a shorter reaction time of 3 h, by modifying the pH of the reaction medium (pH=10 adjusted by $NH_3$ addition). This is based on the literature related to the reduction of GO with endiol structure (HO-A-OH) molecules such as Vitamin C[57], in which the basic conditions favor the hydroxyls dissociation and GO stability, improving the efficiency of the reducing agent.

                14

**Graphene oxide derivates functionalization.** The functionalization of (r)GO with the amino-benzene derivates was carried out using reported methods for ABA[34] and ATP[35]. In the case of the silane functionalization, we established the reaction conditions based on the published literature with GO[36] and colloidal nanocrystals.[58, 59] A GO or rGO dispersion (0.5 mg.mL$^{-1}$) in EtOH was sonicated for 30 min and subsequently functionalized by reflux at 60ºC during 15 h, with 250 μL of MPTS added per mg of (r)GO. The final product was obtained by washing with EtOH to remove the unreacted silane and centrifugation, and finally dispersing the (r)GO in EtOH by sonication (30 min).

**Hybrid material fabrication in solution.** An aliquot of the QD solution in toluene was mixed with a corresponding amount of the silane-*f*-(r)GO dispersion in EtOH at room temperature. The mixture is stirred vigorously during 2 minutes, after which we precipitated the dispersion twice with EtOH to remove the toluene and collect the hybrid material formed. The final product is redispersed in DMF by simply vortexing for at least 12 min. .

**Structural Characterization.** Transmission electron microscopy (TEM) images were acquired with a 100 kV JEOL JEM-1011 microscope. Samples were drop cast onto carbon-coated copper grids for the measurements. X-ray photoelectron spectroscopy (XPS) was carried out on a Kratos Axis Ultra DLD spectrometer, using a monochromatic Al Kα source (15 kV, 20 mA). Samples were prepared by drop casting the materials onto 50 nm-gold coated silicon wafers. High resolution scans were performed at a constant pass energy of 10 eV and steps of 0.1 eV. The photoelectrons were detected at a take-off angle $\Phi = 0º$ with respect to the surface normal. The pressure in the analysis chamber was kept below $7\cdot10^{-9}$ Torr for data acquisition. The binding energy scale was internally referenced to the Au $4f_{7/2}$ peak at 84 eV. Spectra were analysed using CasaXPS software (version 2.3.16). Fitting of S 2p profiles was done assuming Voigt profiles and Shirley-type background. For each S 2p doublet, we fixed a 1.2 eV energy splitting and a 2:1 intensity ratio between the 3/2 and 1/2 components. Each doublet was identified in the text by the position of the $2p_{3/2}$ component.

**Optical Characterization.** Absorption spectra were recorded with a Varian Cary 5000 UV-vis-NIR spectrophotometer. The steady-state and time-resolved photoluminescence (PL) emission were measured using an Edinburgh Instruments FLS920 spectrofluorometer. The steady-state PL

   

was collected exciting the samples at 400 nm. The PL decay traces were recorded exciting the samples at 405 nm with a 50 ps laser diode using a repetition rate of 0.05-1 MHz, to ensure complete decay of the emission between the excitation pulses. The data were collected at the PL peak position with an emission bandwidth of 23 nm.

**Raman Spectroscopy.** Raman data were collected on silicon wafers with a micro-Raman spectrometer Horiba Jobin-Yvon LabRAM HR800UV with a 632.9 nm laser excitation wavelength, using a 50x objective and neutral filter with O.D. 0.6 to reduce the laser intensity. The spectrometer resolution equals 2 cm$^{-1}$. Spectra were recorded after energy scale calibration by checking the Rayleigh and Si bands at 0 and 520.7 cm$^{-1}$. The acquisition time was a few minutes at each point, while the laser excitation power was kept below 1 mW to avoid heating and damage of the sample.

**Conclusions**

In this work we developed a general up-scalable assembly approach from silane-functionalized rGO to prepare solution-processed colloidal QD-rGO dispersions as an *all-in-one* hybrid material, which includes both the absorber and the electron transporting material. Results show the formation of a well-defined hybrid with controlled QDs coverage of the rGO sheets. To demonstrate the potential of the QD-rGO hybrid system for photovoltaic or -detector applications, we have investigated the charge transfer from PbS QDs to rGO by means of fluorescence and Raman spectroscopy. The short MPTS linker leads to efficient electron transfer, evidenced by a significant shortening of the QD lifetime and downshift of the rGO Raman G band. Further studies with core/shell PbS/CdS QDs indicate a barrier effect of the shell on the QD-rGO interactions, which allowed us to modulate the charge transfer by increasing the shell thickness. Therefore, a trade-off between shell thickness and transfer efficiency exists for the device design and performance. Nevertheless, we maintained an efficient transfer for thin shells of 0.2 nm, allowing us to combine the charge transfer with an improved inorganic QD surface passivation to enhance the stability of future QD-rGO hybrid-based devices under ambient conditions.

**Acknowledgements**

We acknowledge support from the European Union 7[th] Framework Programme under grant agreement n° 604391 Graphene Flagship. Francesco De Donato is acknowledged for providing the DIR samples. Riccardo Carzino is acknowledged for technical assistance with the Raman measurements.